\documentclass[12pt]{article} %***

\usepackage[a4paper, total={6.5in, 8.5in}]{geometry}

\usepackage{natbib}

\usepackage{amsmath}
\usepackage{amssymb}
\usepackage{mathabx}

\usepackage{comment}

\usepackage{tikz}
\usetikzlibrary{shapes,snakes}
\usetikzlibrary{positioning}

\usepackage{subfigure}
\usepackage{graphicx}

\usepackage[pdfencoding=auto,
            colorlinks = true,
            urlcolor  = blue,
            ]{hyperref}

\usepackage{multirow}

\usepackage{upgreek}

\usepackage{tablefootnote}

\usepackage{array}
\newcolumntype{C}[1]{>{\centering\let\newline\\\arraybackslash\hspace{0pt}}m{#1}}

\title{Disclosure risk assessment with Bayesian non-parametric hierarchical modelling}

\date{22nd August 2024}
\author{Marco Battiston\\
        \small{School of Mathematical Sciences, Lancaster University}\\
        Lorenzo Rimella\\
        \small{ESOMAS, University of Torino and Collegio Carlo Alberto}\\}

\begin{document}
 \maketitle

%%%%%%%%%%%%%%%%%%%%%%%%%%%%%%%%%%%%%%%%%%%%%%%%%%%%%%%%%%%%%%%%%%%%%%%%%%%%%%%%%%%%%%%%%%%%%%%%%%%%%%%%%%%%%%%%%%%%%%%%%%%%

\abstract{
Micro and survey datasets often contain private information about individuals, like their health status, income or political preferences. Previous studies have shown that, even after data anonymization, a malicious intruder could still be able to identify individuals in the dataset by matching their variables to external information. Disclosure risk measures are statistical measures meant to quantify how big such a risk is for a specific dataset. One of the most common measures is the number of sample unique values that are also population-unique. \cite{Man12} have shown how mixed membership models can provide very accurate estimates of this measure. A limitation of that approach is that the number of extreme profiles has to be chosen by the modeller. In this article, we propose a non-parametric version of the model, based on the Hierarchical Dirichlet Process (HDP).  The proposed approach does not require any tuning parameter or model selection step and provides accurate estimates of the disclosure risk measure, even with samples as small as 1$\%$ of the population size. Moreover, a data augmentation scheme to address the presence of structural zeros is presented. The proposed methodology is tested on a real dataset from the New York census.
}

\section{Introduction} \label{sec1}

Statistical agencies routinely collect and disseminate to the public record-level and micro-data on individual persons and businesses. This data may contain private information about individuals, like their income, political or sexual preferences or health conditions. This creates a serious concern for privacy breaches and the need to protect individuals' anonymity. Previous studies have shown that, even after removing names from the data, a malicious intruder could still be able to identify individuals by matching some of their variables in the dataset to external data. Indeed, in a famous example, \cite{Swe01} was able to identify 97$\%$ of the records in a voter registration list, by using just their birth date and zip code.  

Disclosure risk assessment refers to a broad range of statistical techniques that can be used to assess whether record-level or file-level data has to be considered at risk of disclosing private information. One of the most popular disclosure risk measures,  proposed by \cite{Ski94}, is the number of sample unique values that are also population unique, denoted by $\tau_{1}$. Individuals having a rare or unique combination of values in some variables in the dataset are those most at risk of identification because if their variables are matched using another dataset, it results in a perfect match. Therefore, if the estimated measures of disclosure risk are high, additional privacy-preserving techniques, like for example variable anonymization, data swapping, and addition of noise or cell suppression, should be applied to the dataset before its release to the public.

Among the most popular models to estimate $\tau_{1}$ or similar disclosure risk measures are log-linear, \cite{Ski08}, and mixed-membership (also referred to as grade of membership) models, \cite{Man12}. Log-linear models tend to scale poorly with the number of categorical variables, and their estimates may deteriorate with the presence of many structural zeros.
The mixed membership models, as proposed in 
\cite{Man12}, seem to provide very accurate estimates of $\tau_{1}$, even with samples as small as 1$\%$ or less of the entire population. However, a limitation of this model is the practitioner needs to select a number of extreme profiles $K$ to use for a specific dataset. This has two drawbacks. Firstly, it affects the running time of the methodology, since the model needs to be fitted for different values of $K$ to evaluate differences in the estimates of $\tau_{1}$. Secondly, the choice of a suitable $K$ depends on the value of $\tau_{1}$, which in real data scenarios is not available to the practitioner. 

In this article, we propose a non-parametric version of the mixed-membership model of \cite{Man12} to perform disclosure risk assessment. The proposed model is formulated as a Hierarchical Dirichlet Process, \cite{Teh06}, and allows a potentially unbounded number of extreme profiles. This number is then estimated directly from the data, hence resulting in a tuning-free modelling approach. We describe how to estimate $\tau_{1}$ within the MCMC using both a population sampling approach, as in \cite{Man12}, and a much faster Monte Carlo approximation, which can speed up the computational cost of the algorithm substantially.

A common problem with modelling contingency tables is the presence of many structural zeros. These are combinations of categorical variables that lead to impossible values, i.e. values that are known to be zero in the population, for example, a pregnant male. In real-data applications, structural zeros can account for a very large proportion of possible combinations, and, if their presence is not properly accounted for in the statistical analysis, the performance of the model can deteriorate dramatically. In this article we also describe how to extend the proposed non-parametric model to deal with the presence of many structural zeros, following the data augmentation idea presented in \cite{Man14}.  %Therefore, in Section \ref{sec4}, we describe how to extend the proposed non-parametric model to deal with the presence of many structural zeros, following the data augmentation idea presented in \cite{Man14}. 

To sum up, the paper is organized as follows. Section \ref{sec: review} reviews the disclosure risk problem and introduces the disclosure risk measure $\tau_{1}$. Section \ref{sec3} presents the non-parametric generalization of the mixed membership model of \cite{Man12}, using the Hierarchical Dirichlet Process, and describes the  Markov Chain Monte Carlo (MCMC) algorithm to make inference on the model parameters and to estimate $\tau_{1}$. The extension of the model to include structural zeroes is discussed in Section \ref{sec4}. In Section \ref{sec5}, some empirical illustrations are presented to show the performances of the proposed methodology on synthetic and a real-data example from the New York census data. Finally, some background material and derivations are included in the appendix.

%%%%%%%%%%%%%%%%%%%%%%%%%%%
%%%%%%%%%%%%%%%%%%%%%%%%%%%

\section{Disclosure Risk Problem} \label{sec: review}

Disclosure risk problems for record-level data usually involve two distinct classes of variables: 1) one set of variables, usually called \textit{sensitive variables}, that contain private information, e.g. health status or salary; 2) another class of identifying categorical variables, usually called \textit{key variables}, e.g. gender, age, job, and more general demographic information. Disclosure risk arises because a malicious intruder could potentially identify individuals in the dataset by cross-classifying their key variables and matching them to some external source of information, like publicly available census data. If these matches are correct, the intruder will be able to identify individuals' identities and disclose information contained in their sensitive variables. 
Disclosure risk measures are statistical measures that try to quantify how easy is to identify individuals based on the values of their key variables.

In order to formalize the problem, let us assume that $J$ categorical key variables in the dataset have been observed for a sample of $n$ individuals, sampled from a population of size $N$.  
The $j$-th key variable has $n_{j}$ possible categories, labelled, without loss of generality, from $1$ up to $n_j$. Focusing only on key variables, observation for individual $i$, denoted $X_{i}=\left(X_{i1},\ldots,X_{iJ}\right)$, therefore takes values in the state space $\mathcal{C}:=\bigtimes_{j=1}^{J}\{1\ldots,n_{j}\}$. 
This set has $|\mathcal{C}|=\prod_{j=1}^{J}n_{j}$ values,  corresponding to all possible cross-classification of the $J$ key variables. 
Information about the sample is usually given through the sample frequency vector $\left(f_{1},\ldots,f_{|\mathcal{C}|}\right)$, where $f_{c}$ counts how many individuals out of the $n$ in the sample have a particular combination of cross-classified key variables, corresponding to cell $c\in \mathcal{C}$. $\left(F_{1},\ldots,F_{|\mathcal{C}|}\right)$ denotes the corresponding vector of frequencies in the whole population of $N$ individuals, i.e. $F_{c}$ is the number of individuals in the population belonging to cell $c$.

The earliest papers to consider disclosure risk problems include \cite{Bet90}, \cite{Dun86}, \cite{Dun89}, \cite{Lam93}. These works propose different measures of disclosure risk and ways to estimate them, under specific model assumptions for $\left(f_{1},\ldots,f_{|\mathcal{C}|}\right)$ and $\left(F_{1},\ldots,F_{|\mathcal{C}|}\right)$. \cite{Ski02}, \cite{Ski94} provide reviews of the most popular measures of disclosure risk. Disclosure risk measures depend on the sample frequencies
$\left(f_{1},\ldots,f_{|\mathcal{C}|}\right)$ and often focus on small frequencies, especially on cells having frequency 1, called \emph{sample uniques}. Individuals belonging to these cells are those at the highest risk of having their sensitive information disclosed. This is because if any of these sample unique values are also unique values in the population,  called \emph{population uniques}, any match of their key variables with information from another dataset will produce a perfect match, i.e. perfect certainty about the identity of that specific record, and their sensitive information will be therefore disclosed.  For a review of disclosure risk problems, the reader is referred to \cite{Mat11}.

We usually distinguish between two groups of \emph{measures of disclosure risk}: 
\begin{enumerate}
    \item \textbf{Record-Level} (or per-record) measures: they assign a measure of risk to each data point or specific cell values. Among the most popular ones, there are
   \begin{equation} \label{riskindeces1}
    r_{1c}=\mathbb{P}\left(F_{c}=1|f_{c}=1\right), \ \
        \     r_{2c}=\mathbb{E}\left(1/F_{c}|f_{c}=1\right).
   \end{equation}
   $c\in\{1,\ldots,|\mathcal{C}|\}$.
   %, where $F_{k}$ denotes the number of individuals in the population having the specific combination of key variables corresponding to cell $f_{k}$.
   The first measure provides the probability that a sample unique is also population-unique. The second one gives the probability that, given a sample unique $c$, we guess her identity correctly, by choosing one of the $F_{c}$ values in the population uniformly at random. In general, the first measure is less conservative and is always smaller than the second. 
    \item \textbf{File-level} measures: they provide an overall measure of risk for an entire sample or dataset. File-level measures are usually defined by aggregating the corresponding record-level ones. Popular examples are
    \begin{equation} \label{riskindeces2}
        \tau_{1}=\sum_{c\in \mathcal{C}:f_{c}=1}r_{1c}, \ \ \
        \tau_{2}=\sum_{c\in \mathcal{C}:f_{c}=1}r_{2c}.
    \end{equation}
\end{enumerate}

In the disclosure risk literature, $\tau_{1}$ is the most popular measure of disclosure risk and, in the rest of the paper, we will focus on its estimation using the data $\left(f_{1},\ldots,f_{|\mathcal{C}|}\right)$. In the literature, the most popular modelling choices for this task are   
\emph{log-linear} and \emph{mixed membership} models. Regarding the former ones, the main references are \cite{Ski08}, \cite{Shl10}, in which indexes \eqref{riskindeces1} and \eqref{riskindeces2} are derived in closed form and estimated using plug-in MLE estimators. Regarding the latter class of models, \cite{Man12} proposed the use of mixed membership models, which resulted in very accurate estimates for \eqref{riskindeces2}, even for sample sizes $n$ much smaller than the population size $N$.

If the estimated values of \eqref{riskindeces1} and \eqref{riskindeces2} are too high, then the data curator should apply a disclosure limitation technique to the dataset, before releasing it to the public. Some possibilities are for example rounding, data swapping, cell suppression of extreme values or entire variables, subsampling or perturbation techniques. See \cite{Wil01} for a review of different disclosure limitation techniques.

%\section{Mixed Membership models for disclosure risk assessment} \label{sec3}
\section{Mixed Membership models} \label{sec3}

In this section, we extend the mixed membership model of \cite{Man12}, reviewed in the appendix, to its non-parametric version. Then, we summarize both the MCMC sampler to perform posterior inference and describe how to estimate \eqref{riskindeces2} within the sampler using either population sampling, as in \cite{Man12}, or a faster Monte Carlo approximation. 

%In this section, we firstly briefly review the mixed membership model of \cite{Man12} in subsection \ref{sec:ModelMan}. Then, we describe extend it to a non-parametric version in subsection \ref{sec:ModelHDP}. Finally, \ref{sec:sampling}, we summarize both the MCMC sampler to perform posterior inference and describe how to estimate \eqref{riskindeces2} within the sampler using either population sampling, as in \cite{Man12}, or a faster Monte Carlo approximation. \\
In terms of background about the Hierarchical Dirichlet Process and its properties (Stick Breaking construction, Chinese Restaurant Franchise, Posterior representation), the reader is referred to \cite{Teh10}

\subsection{Non-parametric Mixed Membership Model} \label{sec:ModelHDP}

\emph{Mixed Membership models} are generalizations of mixture models to model multiple groups of observations. In their parametric version, they assume $K$ extreme profiles (alias mixture components), having weights in the population regulated by a $K$-dimensional probability vector $\pmb{g}_{0}$. Within each group, some heterogeneity from the common proportions $\pmb{g}_{0}$ is allowed, by introducing a group-specific partial affiliation vector $\pmb{g}_{i}$. In the model used in \cite{Man12}, the $i$-th group of observations corresponds to the $J$ observations of key variables of the $i$-th individual.

In order to allow an unbounded number of extreme profiles, we select $G_{0} \sim \text{DP}\left(\alpha_{0},H\right)$, where $\text{DP}$ is a Dirichlet Process, \cite{Fer73}, with concentration parameter $\alpha_{0}$ and base measure $H$. The base measure $H$ is a probability measure on the space of all arrays with $J$ rows, having a $n_{j}$-dimensional probability vector in the $j$-th row. From the stick-breaking representation of the Dirichlet Process, $G_{0}$ can be represented as
\begin{equation*}
    G_{0} = \sum_{k=1}^{\infty}g_{0,k}\delta_{\theta^{\left(k\right)}}
\end{equation*}
$\left(\theta^{\left(k\right)}\right)_{k=1}^{\infty}$ are independent and identically distributed arrays, sampled from the base measure $H$, representing the likelihood of the possibly unbounded extreme profiles, while the sequence $\left(g_{0,k}\right)_{k=1}^{\infty}$ is such that all entries $0\leq g_{0,k}\leq 1$ and $\sum_{k}g_{0,k}=1$, and is sample following a stick breaking distribution of parameter $\alpha_{0}$, \cite{Set94}. As in the parametric case, $g_{0,k}$ can be thought of as the popularity of extreme profile $\theta^{\left(k\right)}$ in the population. 

Given $G_{0}$, each individual $i$ selects her own affiliation distribution $G_{i}$, representing her partial affiliation to each possible extreme profile, according to $G_{i}|G_{0} \sim \text{DP}\left(\alpha_{i},G_{0}\right)$. Given the almost sure discreteness of $G_{0}$, each $G_{i}$ is supported on the same atoms of $G_{0}$ and can be represented as
\begin{align*}
    G_{i} = \sum_{k=1}^{\infty} g_{i,k}\delta_{\theta^{\left(k\right)}}
\end{align*}
for a sequence of probability weights $\left(g_{i,k}\right)_{k=1}^{\infty}$, see pages 161-162 of \cite{Teh10}. The parameter $\alpha_{i}$ regulates the variability of the weights $\left(g_{i,k}\right)_{k=1}^{\infty}$ around their mean value $\left(g_{0,k}\right)_{k=1}^{\infty}$. The higher $\alpha_{i}$, the more heterogeneous individual $i$ is from the rest of the population.

Given the individual specific affiliation vector $G_{i} $, individual $i$ will select her $j$-th key variable from the infinite mixture model 
\begin{equation*} 
 X_{i,j}|G_{i} \sim \sum_{k=1}^{\infty}g_{i,k}\theta^{\left(k\right)}_{j,\cdot} 
\end{equation*}
where $\theta^{\left(k\right)}_{j,\cdot}$ denotes the $j$-th row of $\theta^{(k)}$.

As in the finite-dimensional case, it is computationally convenient to  introduce the mixture classification variables $Z_{i,j}$, taking integer values, and summarize the model as follows,
\begin{align*}   X_{i,j}|\left(Z_{i,j}=k\right),\left(\theta_{k}\right)_{k=1}^{\infty}  & \sim \theta^{\left(k\right)}_{j,\cdot} \ \ \ i=1,\ldots,n, \ \ \ j=1,\ldots,J \\     \mathbb{P}\left(Z_{i,j}=k|G_{i}\right) &= g_{i,k} \quad k\in \mathbb{N}, \quad  i=1,...,n, \quad j=1,...,J  \\
    G_{i}|\alpha,G_{0} &\sim \text{DP}\left(\alpha_{i},G_{0}\right) \ \ \  i=1,\ldots,n \\
    G_{0} &\sim \text{DP}\left(\alpha_{0},H\right) \\
    \alpha_{i} &\sim \text{Ga}\left(a,b\right) \ \ \  i=1,\ldots,n  \nonumber\\
   \alpha_0 &\sim \text{Ga}\left(a_0,b_0\right).
\end{align*}
where the base measure $H$ is chosen to assign $\text{Dir}\left( \mathbb{I}_{n_{j}}\right)$ prior to the $j$-th row, for each $j\in\{1,\ldots,J\}$, where $\mathbb{I}_{n_{j}}$ is a vector of dimension $n_{j}$ with all entries equal to $1$. Finally, $\text{Ga}$ denotes a Gamma distribution, and $a,b,a_{0},b_{0}$ are positive hyperparameters.

%%%%%%%%%%%%%%%%%%%%%%%%%%%%%%
%%%%%%%%%%%%%%%%%%%%%%%%%%%%%%
%%%%%%%%%%%%%%%%%%%%%%%%%%%%%%

\subsection{Posterior Inference} \label{sec:sampling}

\subsubsection{MCMC sampler}
\label{sec: direct.assign}
Posterior inference of the model parameters can be performed using the Direct Assignment algorithm for the Hierarchical Dirichlet Process, pages 196-199 of \cite{Teh10}. In the sampler, $m_{ik}$ denotes the number of tables in individual $i$ assigned to mixture component $k$. At any stage of the algorithm, we denote by $K_{n}$ the number of active mixture components, i.e. components $\theta^{\left(k\right)}$ with at least one $Z_{i,j}$ assigned to them. At step 3, the sampler resamples $\left(g_{0,k}\right)_{k=1}^{\infty}$, by drawing a probability vector $\left(g_{0,k}\right)_{k=0}^{K_{n}}$ (using the posterior representation of $G_{0}$, formula 5.9 in \cite{Teh10}), where $g_{0,0}$ represent the probability of a new mixture, i.e. $g_{0,0}=1-\sum_{k=1}^{K_{n}}g_{0,k}$. Similarly for $\left(g_{i,k}\right)_{k=1}^{\infty}$ at step 4. For ease of notation, we will simply write $\left(g_{0,k}\right)$ and $\left(g_{i,k}\right)$, where the index is over $k$ and ranges from $0$ to $K_{n}$. \\ 
The sampler iterates over the following steps,

\begin{enumerate}
    \item  Sample $Z_{i,j}$: for $i \in \{1,\ldots,n\}$ and $j \in \{1,\ldots, J \}$, sample $Z_{i,j}$ from 
     \begin{align} \label{eq:fc.Z_ij}
            Z_{i,j} = \begin{cases} k & \text{with prob} \ \ \propto \ \ g_{i,k} \theta^{\left(k\right)}_{j,x_{ij}}   \\
         k^{\text{new}} &  \text{with prob} \ \ \propto 
          \ \ g_{i,0}\frac{1}{n_{j}} \end{cases}
        \end{align}
      for   $k\in \{1,\ldots,K_{n}\}$, where the factor $1/n_{j}$ is the marginal probability of $Z_{i,j}$ being sampled from a new mixture $\theta^{\left(K_{n}+1\right)}$, when $\theta^{\left(K_{n}+1\right)}$ is distributed according to $H$. \\
    If $Z_{i,j}= k^{\text{new}}$, draw $\theta^{\left(k^{\text{new}}\right)}$ from \eqref{eq:post.thetak}, and update $\left(g_{0,k}\right)$ and $\left(g_{i,k}\right)$ as follows
    \begin{align*}
        \nu_{0} | \alpha_{0} & \sim \text{Beta}\left(\alpha_{0},1\right) \\
\left(g_{0,0}^{\text{new}},g^{\text{new}}_{0,K_n+1}\right) & = \left(g_{0,0}\nu_{0},g_{0,0}\left(1-\nu_{0}\right) \right) \\
    \nu_{i}|g_{0,0},\alpha,\nu_{0} & \sim \text{Beta}\left(\alpha g_{0,0}\nu_{0},\alpha g_{0,0}\left(1-\nu_{0}\right)\right) \\
\left(g_{i,0}^{\text{new}},g_{i,K_n+1}^{\text{new}}\right) & = \left(g_{i,0}\nu_{i},g_{i,0}\left(1-\nu_{i}\right)\right)
    \end{align*}
    for every $i=1,\ldots,n$. Finally, set $Z_{i,j}=K_{n}+1$ and increment $K_{n}$ by $1$.

    \item Sample $m_{ik}$: for $i\in\{1,\ldots,n\}$ and $k\in\{1,\ldots,K_{n}\}$, compute $n_{i\cdot k} = \sum_{j=1}^{J}\mathbb{I}\left(Z_{i,j}=k\right)$, and sample $m_{ik}$ from
    \begin{equation*}
        \mathbb{P}\left(m_{ik}=m| n_{i\cdot k} ,g_{0,k},\alpha_{0} \right) = \frac{\Gamma\left(\alpha_{0}g_{0,k} \right) }{\Gamma\left(\alpha_{0}g_{0,k} +n_{i\cdot k} \right)} s\left(n_{i\cdot k} ,m\right)\left(\alpha_{0}g_{0,k}\right)^{m}
    \end{equation*}
    for $m\in\{1,\ldots,n_{i\cdot k}\}$, and where $s\left(n,m\right)$ are the unsigned Stirling numbers of the first kind, which can be pre-computed outside the sampler from the recursion, $s\left(0,0\right)=s\left(1,1\right)=1$, $s\left(n,0\right)=0$ for $n>0$ and $s\left(n,m\right)=0$ for $m>n$ and $s\left(n+1,m\right)=s\left(n,m-1\right)+n s\left(n,m\right)$. \\
    As an alternative, $m_{ik}$ can also be computed by drawing a Chinese Restaurant Process with $n_{i\cdot k}$ customers and concentration parameter $\alpha_{0}g_{0,k}$, and setting $m_{ik}$ equal to the number of resulting tables. This approach is very fast when the number $J$ of categorical variables is small.
    \item Sample $\left(g_{0,k}\right)$: compute $m_{\cdot k} = \sum_{i=1}^{n}m_{ik}$ for $k=1,\ldots,K_{n}$, and resample $\left(g_{0,k}\right)$ from
    \begin{equation} \label{eq:fc.beta}
        \text{Dir}\left(\alpha_{0},m_{\cdot 1}, \ldots, m_{\cdot K_{n}}\right)
    \end{equation}
    \item Sample $\left(g_{i,k}\right)$: for $i\in \{1,\ldots, n\}$, resample $\left(g_{i,k}\right)$ from
    \begin{equation} \label{eq:fc.pi_i}
        \text{Dir}\left(\alpha_{i}g_{0,0},\alpha_{i}g_{0,1} + n_{i\cdot 1}, \ldots,\alpha_{i}g_{0,K_{n}} + n_{i\cdot K}\right)
    \end{equation}
    \item Sample $\theta^{\left(k\right)}$: for $k\in\{1,\ldots,K_{n}\}$ and $j\in \{1,\ldots,J\}$ sample $\theta_{j,\cdot}^{\left(k\right)}$ according to, % \eqref{theta},
    \begin{equation} \label{eq:post.thetak}
         \text{Dir}\left(1+\sum_{i=1}^n \mathbb{I}\left(Z_{ij} = k, X_{ij} = 1\right),...,1+\sum_{i=1}^n \mathbb{I}\left(Z_{ij} = k, X_{ij} = n_j\right)\right)
    \end{equation}
\item  Sample $\alpha_{0},\alpha_{i}$: Using the augmentation from the Appendix of \cite{Teh06}, let $m_{\cdot \cdot} = \sum_{k=1}^{K_{n}} m_{\cdot k}$, then sample $\alpha_0$ according to
\begin{align*}
    \eta_0|\alpha_0,m_{\cdot \cdot} &\sim \text{Beta}\left(\alpha_0 + 1, m_{\cdot \cdot}\right) \\
    s_{0}|m_{\cdot\cdot},\eta_{0},K_{n} & \sim \text{Bern}\left( \frac{m_{\cdot\cdot}\left(b_{0}-\log \eta_{0}\right) }{K_{n}+a_{0}-1+m_{\cdot\cdot}\left(b_{0}-\log \eta_{0}\right) }\right) \\
     \alpha_0|\eta_0,s_{0},K_{n} &\sim \text{Gamma}\left(a_0+K_{n}-s_{0}, b_0-\log\eta_0\right)
\end{align*}
and $\alpha_{i}$, for $i\in\{1,\ldots,n\}$ according to
\begin{align*}
    \eta_i|\alpha_{i}, J &\sim \text{Beta}(\alpha_{i} + 1, J) \quad i=1,...,n; \\
     s_{i}|m_{i\cdot},\eta_{i} & \sim \text{Bern}\left( \frac{J(b-\log \eta_{i}) }{m_{i\cdot}+a-1+J(b-\log \eta_{i}) }\right) \quad i=1,...,n, \\
   \alpha_{i}|\eta_i, m_{i \cdot} &\sim \text{Gamma}(a+m_{i \cdot} -s_{i} , b-  \log \eta_i) \quad i=1,...,n. 
\end{align*}

\end{enumerate}

\subsubsection{Estimation of $\tau_{1}$}
\label{sec: estim.tau}

In this section, we describe two approaches to estimate the disclosure risk measure $\tau_{1}$, formula \eqref{riskindeces2}, within the sampler described in \ref{sec:sampling}. 

The first approach follows \cite{Man12} and relies on the simulation of the unobserved individuals in the population. Specifically, remember that $f=\left(f_{1},\ldots,f_{|\mathcal{C}|
}\right)$ denotes the vector of frequencies of each cell $c\in \mathcal{C}$ in the sample of size $n$, and $F$ the corresponding vector in the population of size $N$. Then, at iteration $m$ of the MCMC sampler, the $m$-th draw of $\tau_{1}^{\left(m\right)}$ can be obtained by applying the following algorithm
\begin{enumerate}
    \item Let $F^{\left(m\right)}=f$, i.e. initialize the population vector using the sample vector;
    \item for $i=n+1,\ldots,N$: 
    \begin{enumerate}
        \item Draw $\left(g_{i,k}\right)$ from \eqref{eq:fc.pi_i};
        \item For $j=1,\ldots,J$: 
        \begin{enumerate}
            \item Sample $Z_{i,j}|\left(g_{i,k}\right)$ from \eqref{eq:fc.Z_ij};
            \item Sample $X_{i,j}\sim \theta_{j,\cdot}^{\left(Z_{i,j}\right)}$;
        \end{enumerate}
        \item Set $F^{\left(m\right)}_{c}=F^{\left(m\right)}_{c}+1$, where $c$ is the cell corresponding to the sampled $X_{i}$;
    \end{enumerate}
    \item Set $\tau_{1}^{\left(m\right)}=\sum_{c\in \mathcal{C}}\mathbb{I} \left(F_{c}^{m}=1,f_{c}=1\right)$, where $\mathbb{I}$ denotes the indicator function.
\end{enumerate}
Point estimates and credible intervals of $\tau_{1}$ can then be obtained from the empirical quantities. This approach is computationally intensive when the population $N$ is large.

An alternative approach, computationally much faster, relies on a Monte Carlo approximation. Specifically,
let us recall that $\mathcal{C}:=\bigtimes_{j=1}^{J}\{1\ldots,n_{j}\}$ denotes the state space of the observations. Given a sample $X_{1:n}:=\left(X_{1},\ldots,X_{n}\right)$, and denoting by $\tilde{\mathcal{C}}_{X_{1:n}}$ the set $\tilde{\mathcal{C}}_{X_{1:n}}:=\{c\in \mathcal{C}: \sum_{c\in\mathcal{C}} \mathbb{I}\left(\sum_{i=1}^{n}\mathbb{I}\left(X_{i}=c\right)=1\right)\}$ the set of combinations appearing with frequency $1$ in the sample. Then, $\tau_{1}$ can be estimated within the MCMC using the following algorithm,

\begin{enumerate}
\item For $t =1,\ldots T$:  draw $\left(g_{t,k}\right)$ from \eqref{eq:fc.pi_i};
    \item For $c\in \tilde{\mathcal{C}}_{X_{1:n}}$: 
     Compute the Monte Carlo approximation,
        \begin{equation*}
            \mathbb{P}\left( \{X_{n+1}= c \} |G_{0},\alpha_{0}\right) \approx \frac{1}{T}\sum_{t=1}^{T}\prod_{j=1}^{J} \left(\sum_{k=1}^{K_{n}} g_{t,k}  \theta^{\left(k\right)}_{j,c_j}  + g_{t,0} \frac{1}{n_{j}}\right)
        \end{equation*}
    \item Set $\tau_{1}^{\left(m\right)}=\sum_{c\in\tilde{\mathcal{C}}_{X_{1:n}}}  \left(1- \mathbb{P}\left( \{X_{n+1}= c \} |G_{0},\alpha_{0}\right)\right)^{N-n}$.
\end{enumerate}

% \begin{enumerate}
%     \item For $c\in \tilde{\mathcal{C}}_{X_{1:n}}$: 
%     \begin{enumerate}
%         \item For $t =1,\ldots T$:  draw $\left(g_{t,k}\right)$ given $\left(g_{0,k}\right)$ from \eqref{eq:fc.pi_i};
%         \item Compute the Monte Carlo approximation,
%         \begin{equation*}
%             \mathbb{P}\left( \{X_{n+1}= c \} |G_{0},\alpha_{0}\right) \approx \frac{1}{T}\sum_{t=1}^{T}\prod_{j=1}^{J} \left(\sum_{k=1}^{K_{n}} g_{t,k}  \theta^{\left(k\right)}_{j,c_j}  + g_{t,0} \frac{1}{n_{j}}\right)
%         \end{equation*}
%     \end{enumerate}
%     \item Set $\tau_{1}^{\left(m\right)}=\sum_{c\in\tilde{\mathcal{C}}_{X_{1:n}}}  \left(1- \mathbb{P}\left( \{X_{n+1}= c \} |G_{0},\alpha_{0}\right)\right)^{N-n}$.
% \end{enumerate}

In the algorithm, $T$ is the number of Monte Carlo draws to approximate $\mathbb{P}\left( \{X_{n+1}= c \} |G_{0},\alpha_{0}\right)$. The algorithm is easily parallelizable both in $t$ and $c$. The derivations of this approximation, and the corresponding formula for the algorithm with structural zeros of Section \ref{sec4}, can be found in the appendix. 

% \textcolor{red}{
% In the code, are we doing as above or ?
% \begin{enumerate}
% \item For $t =1,\ldots T$:  draw $\left(g_{t,k}\right)$ from \eqref{eq:fc.pi_i};
%     \item For $c\in \tilde{\mathcal{C}}_{X_{1:n}}$: 
%      Compute the Monte Carlo approximation,
%         \begin{equation*}
%             \mathbb{P}\left( \{X_{n+1}= c \} |G_{0},\alpha_{0}\right) \approx \frac{1}{T}\sum_{t=1}^{T}\prod_{j=1}^{J} \left(\sum_{k=1}^{K_{n}} g_{t,k}  \theta^{\left(k\right)}_{j,c_j}  + g_{t,0} \frac{1}{n_{j}}\right)
%         \end{equation*}
%     \item Set $\tau_{1}^{\left(m\right)}=\sum_{c\in\tilde{\mathcal{C}}_{X_{1:n}}}  \left(1- \mathbb{P}\left( \{X_{n+1}= c \} |G_{0},\alpha_{0}\right)\right)^{N-n}$.
% \end{enumerate}
% }

%%%%%%%%%%%%%%%%%%%%%%%%%%%%%%
%%%%%%%%%%%%%%%%%%%%%%%%%%%%%%
%%%%%%%%%%%%%%%%%%%%%%%%%%%%%%

\section{Extension to Structural Zeros} \label{sec4}

Structural zeros are combinations of key variables that lead to impossible values, like a five-year-old veteran or a pregnant male. In real datasets, structural zeros might account for an extremely large proportion of the possible cells $|\mathcal{C}|$, often above 90-95$\%$. If a Bayesian model does not take into account the presence of structural zeros, its posterior estimates can deteriorate dramatically, as shown in an example in Section \ref{sec5}. This is because, if the prior distribution assigns positive probability to every possible cell in $\mathcal{C}$, the posterior distribution will also assign some mass to every cell. Even if the posterior mass assigned to each structural zero cell is very low, if the number of these cells is very large, their overall posterior mass will be far from being negligible. 

Following the general algorithm of \cite{Man14}, in this Section, we describe an MCMC algorithm to perform posterior inference on the model parameters and disclosure risk measure $\tau_{1}$, in the presence of structural zeros. 
The main idea of the algorithm is to consider the observed sample, $X_{1:n}$ of size $n$, as a truncated version of larger sample $X_{1:n+n_0}$ of size $n+n_{0}$ sampled from the model of section \ref{sec:ModelHDP}, and in which $n$ of these observations have fallen into admissible cells, while the other $n_{0}$ have taken values in structural zeros cells. Then, the algorithm is a data augmentation scheme in which in steps 7-9, we sample the latent variables ($n_{0}$ truncated observations in the structural zeros, denoted $Z_{\left(n+1\right):\left(n+n_{0}\right)}$ $X_{\left(n+1\right):\left(n+n_{0}\right)}$) given the observed variables and common parameters, and in steps 3-5 we sample the common parameters given both the observed and latent variables. 

Structural zeros can be defined in terms of marginal conditions. These are conditions that fix 2 or more key variables to some specific values. For example $\mu =\{ *, 1, *, *, 2, *\}$ is the marginal condition on a dataset with 6 key variables and includes all cells taking value 1 in the second variable and value 2 in the fifth one, and the placeholder symbol $*$ means that that variable is unrestricted. Conditions that fix more than one category in a specific variable can be written separately as unions of multiple marginal conditions. Moreover, a set of overlapping marginal conditions can always be rewritten as a (possibly larger) set \emph{disjoint} marginal conditions. For example, let us suppose to have 3 binary key variables, the two overlapping conditions $\tilde{\mu}_{1} =\{ *, 1, 2\}$ and $\tilde{\mu}_{2} =\{ 1, 1, *\}$ (cell $\{1,1,2\}$ belongs to both conditions) can be rewritten as disjoint conditions $\mu_{1} =\{ *, 1, 2\}$ and $\mu_{2} =\{ 1, 1, 1\}$, i.e. $\tilde{\mu}_{1}\cup\tilde{\mu}_{2}=\mu_{1}\cup\mu_{2}$.

Section 4.2 of \cite{Man14} presents a simple algorithm to transform a set of \emph{overlapping} marginal conditions into a set of \emph{disjoint} ones. This algorithm is run as a pre-processing step before implementing the MCMC. Therefore, we can assume to have a set of $C$ \emph{disjoint} marginal conditions, denoted $S_{d}=\{ \mu_{1},\ldots, \mu_{C} \}$, specifying sets of impossible cells, and $S=\cup_{c=1}^{C}\mu_{c}$ the subset of sample space $\mathcal{C}$ corresponding to structural zeros. In the MCMC sampler, for each marginal constraint $\mu_{c}$, steps 7-9 simulate the truncated observations from $X_{\left(n+1\right):\left(n+n_{0}\right)}$ that fall into the cells specified by $\mu_{c}$, and their corresponding mixture classification variables $Z_{\left(n+1\right):\left(n+n_{0}\right)}$. Specifically, step 7 computes the probability $p_{c}$ of all cells in $\mu_{c}$, step 8 samples the number $n_{c}$ of truncated observations from $X_{\left(n+1\right):\left(n+n_{0}\right)}$ in $\mu_{c}$, and finally step 9 samples their mixture classification $Z_{i}$ given the event $X_{i} \in \mu_{c}$.

%%%%%%%%

\subsection{MCMC Algorithm including structural zeros} \label{sec:mcmc_struc}

The MCMC sampler of Section \ref{sec: direct.assign} can be extended following \cite{Man14} to account for the presence of structural zeros. Specifically, we repeat the following steps

\begin{enumerate}
    \item  Sample $Z_{i,j}$: for $i \in \{1,\ldots,n\}$ and $j \in \{1,\ldots, J \}$, sample $Z_{i,j}$ from \eqref{eq:fc.Z_ij}.

    \item Sample $m_{ik}$: for $i=\{1,\ldots,n + n_{0}\}$ and $k=\{1,\ldots,K_{n}\}$, sample $m_{ik}$ as in step 2 of the sampler in Section \ref{sec:sampling}.
    
    \item Draw $\left(g_{0,k}\right)$: Compute $m_{\cdot k} = \sum_{i=1}^{n+n_{0}}m_{ik}$ for $k\in \{1,\ldots,K_{n}\}$, and sample $\left(g_{0,k}\right)$ from \eqref{eq:fc.beta}.  
    
    \item Draw $\left(g_{i,k}\right)$: for $i\in \{1,\ldots, n\}$ sample $\left(g_{i,k}\right)$ from \eqref{eq:fc.pi_i}.

    \item Draw $\theta^{\left(k\right)}$: For $k\in\{1,\ldots,K_{n}\}$ and $j\in \{1,\ldots,J\}$ sample $\theta_{j,\cdot}^{\left(k\right)}$ according to, % \eqref{theta},
    \begin{equation*} 
         \text{ Dirichlet}\left(1+\sum_{i=1}^{n+n_{0}} \mathbb{I}\left(Z_{ij} = k, X_{ij} = 1\right),...,1+\sum_{i=1}^{n+n_{0}} \mathbb{I}\left(Z_{ij} = k, X_{ij} = n_j\right)\right)
    \end{equation*}

\item  Update $\alpha_{0},\alpha_{i}$, for $i\in\{1,\ldots,n\}$, as in step 6 of the sampler in Section \ref{sec:sampling}.

\item Compute $\left(p_{1},\ldots,p_{C}\right)$: for $c \in \{1,\ldots,C\}$, compute with Monte Carlo
\begin{align*}
    p_{c} := \mathbb{P} & \left( X_{i}\in \mu_{c} |G_{0},\alpha\right)  = \int \mathbb{P}\left(X_{i}\in \mu_{c} |G_{i},\alpha\right)\mathbb{P}\left(G_{i}|G_{0},\alpha\right) dP\left(G_{i}\right) \\
  &=  \int  \prod_{j\in \{1,\ldots,J\}:\mu_{c,j}\neq  *} \left(\sum_{k=1}^{K_{n}} g_{i,k}  \theta^{\left(k\right)}_{j,\mu_{c,j}}  + g_{i,0} \frac{1}{n_{j\cdot\cdot}}\right) \text{Dir}\left((g_{i,k})|(g_{0,k})\right)dg_{i,k}
\end{align*}

\item Draw $\left(n_{1},\ldots,n_{C}\right)$: sample a vector
     \begin{equation*}
         \left(n_{1},\ldots,n_{C}\right)\sim \text{NM}\left(n,p_{1},\ldots,p_{C}\right),
     \end{equation*}
     where $\text{NM}$ denotes a Negative Multinomial distribution, with mass function,
     \begin{equation*}
p\left(n_{1},\ldots,n_{C}|p_{1},\ldots,p_{C}\right)= \frac{\Gamma\left(n+n_{0}\right)}{\Gamma\left(n\right)\prod_{c=1}^{C}n_{c}!} \left(1 - \sum_{c=1}^{C}p_{c}\right)^{n} \prod_{c=1}^{C}p_{c}^{n_{c}}
     \end{equation*}
     where $n_{0} :=\sum_{i=1}^{C}n_{i}$.

\item Sample $Z_{\left(n+1\right):\left(n+n_{0}\right)}$, $X_{\left(n+1\right):\left(n+n_{0}\right)}$: for $c \in \{1,\ldots,C\}$ and for $i \in \{ n + \sum_{l=1}^{c-1}n_{l}+1, n + \sum_{l=1}^{c-1}n_{l}+ 2, \ldots, n + \sum_{l=1}^{c-1}n_{l}+n_{c} \}$ (with the proviso that $\sum_{l=1}^{0}n_{l}=0$):
\begin{itemize}
    \item Draw $\left(g_{i,k}\right)$ from \eqref{eq:fc.pi_i}. Then, for $j \in \{1,\ldots,J\}$:
    \begin{itemize}
    \item If $\mu_{c,j}\neq *$: Set $X_{i,j}=\mu_{c,j}$ and sample $Z_{i,j}$ from \eqref{eq:fc.Z_ij}.
    \item If $\mu_{c,j} = *$: Sample $\mathbb{P}\left(Z_{i,j}=k|\left(g_{i,k}\right)\right) =g_{i,k}$, and sample $X_{i,j}|Z_{i,j},\theta \sim \theta^{\left(Z_{i,j}\right)}_{j,\cdot}$.
\end{itemize}
\end{itemize}
\end{enumerate}

When the structural zeros account for the majority of cells, the probabilities and counts from steps 7 and 8 can become very large. This implies that, at step 9, many variables have to be simulated, and this slows down the algorithm significantly. In the appendix, we describe an approximation of step 9 that reduces the computational cost dramatically and produces similar estimates of $\tau_{1}$. Remark also that even the estimate of $\tau_1$ should account for the presence of structural zeros, and, as already mentioned, we describe this modification in the appendix.

%In order to estimate $\tau_{1}$,

%%%%%%%%%%%%%%%%%%%%%%%%%%%%%%%%%%%%%%
%%%%%%%%%%%%%%%%%%%%%%%%%%%%%%%%%%%%%%
%%%%%%%%%%%%%%%%%%%%%%%%%%%%%%%%%%%%%%

\section{Experiments} \label{sec5}

This section is composed of two parts. In Subsection \ref{sec:exp.synt}, we compare the parametric and non-parametric versions of the mixed membership model on synthetic data. In Subsection \ref{sec:exp.real}, we test the performances of the non-parametric model on a real dataset in two scenarios, with and without modelling the structural zeros.

The code to reproduce the experiments is open source and available at \\\href{https://github.com/LorenzoRimella/BNP_DR}{https://github.com/LorenzoRimella/BNP\_DR}. All the experiments were run on a 32 GB Tesla V100 GPU available on ``The High-End Computing'' (HEC) facility at Lancaster University. In the appendix, we have also included some notes on how we implement the algorithm to exploit parallel computing.

\subsection{Synthetic data} \label{sec:exp.synt}

We generated synthetic data of size $N=712174$ from the mixed membership models of \cite{Man12}, for $J=10$ categorical variables, and different number of categories $n_{j}$ per variable, ranging from $2$ to $11$. We draw three samples of sizes $n=1000,5000,10000$, and run three MCMC samplers: i) the non-parametric HDP model with $\tau_{1}$ estimated via population sampling; ii) the non-parametric HDP model with $\tau_{1}$ estimated via Monte Carlo sampling; iii) the algorithm from \cite{Man12} for different values of $K$, for 400k iterations, out of which 300k discarded as a burn-in. 

\begin{figure}[httb!]
    \centering
    \includegraphics[width=0.8\textwidth]{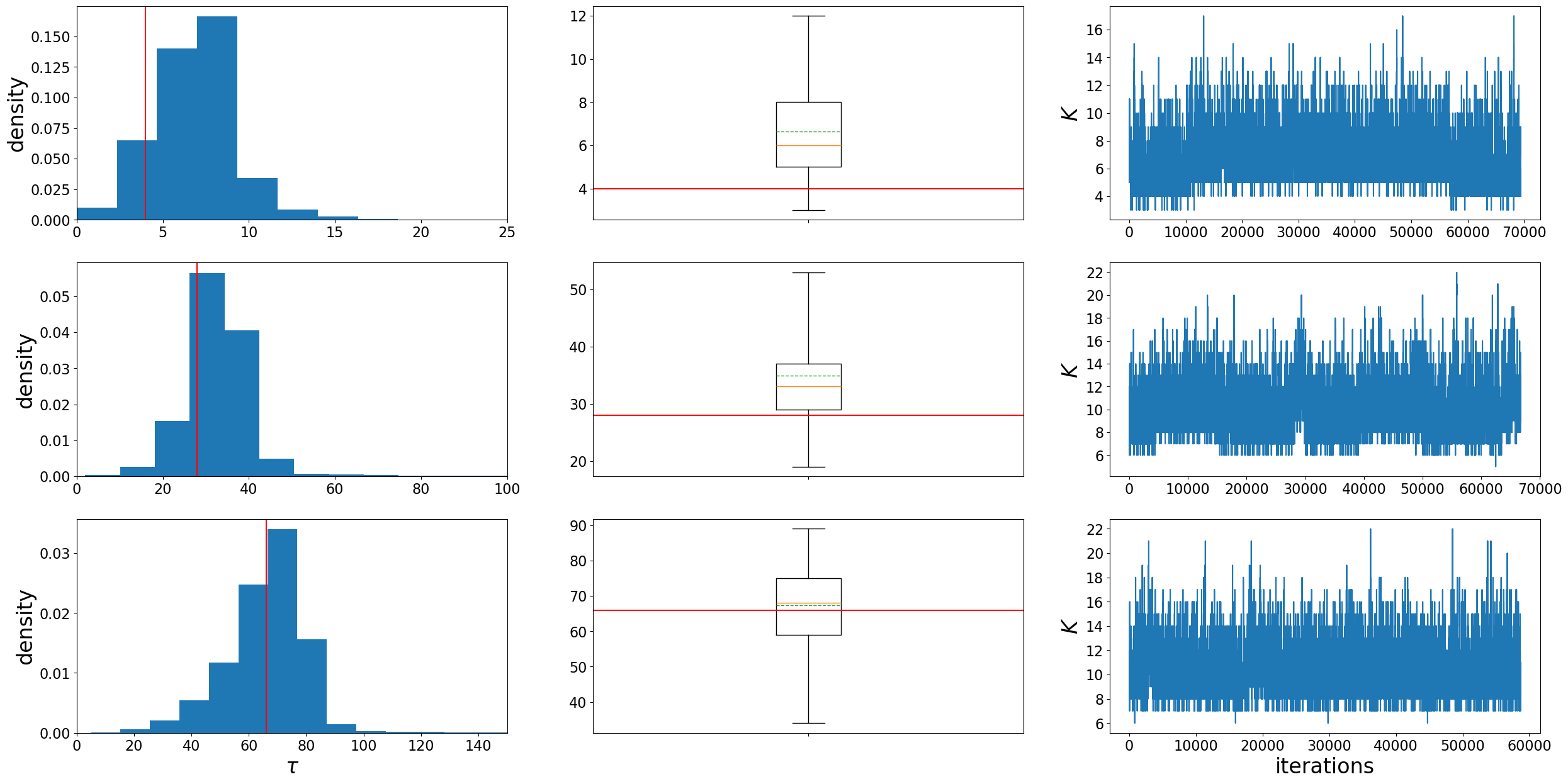}
    \caption{HDP on synthetic data with $\tau_{1}$ estimated via sampling. In red the true $\tau_1$. For the box plots: in orange the median, in dashed green the mean, and the whiskers show 95\% credible intervals.}
    \label{fig:synth_BNP_sampling}
\end{figure}

\begin{figure}[httb!]
    \centering
    \includegraphics[width=0.8\textwidth]{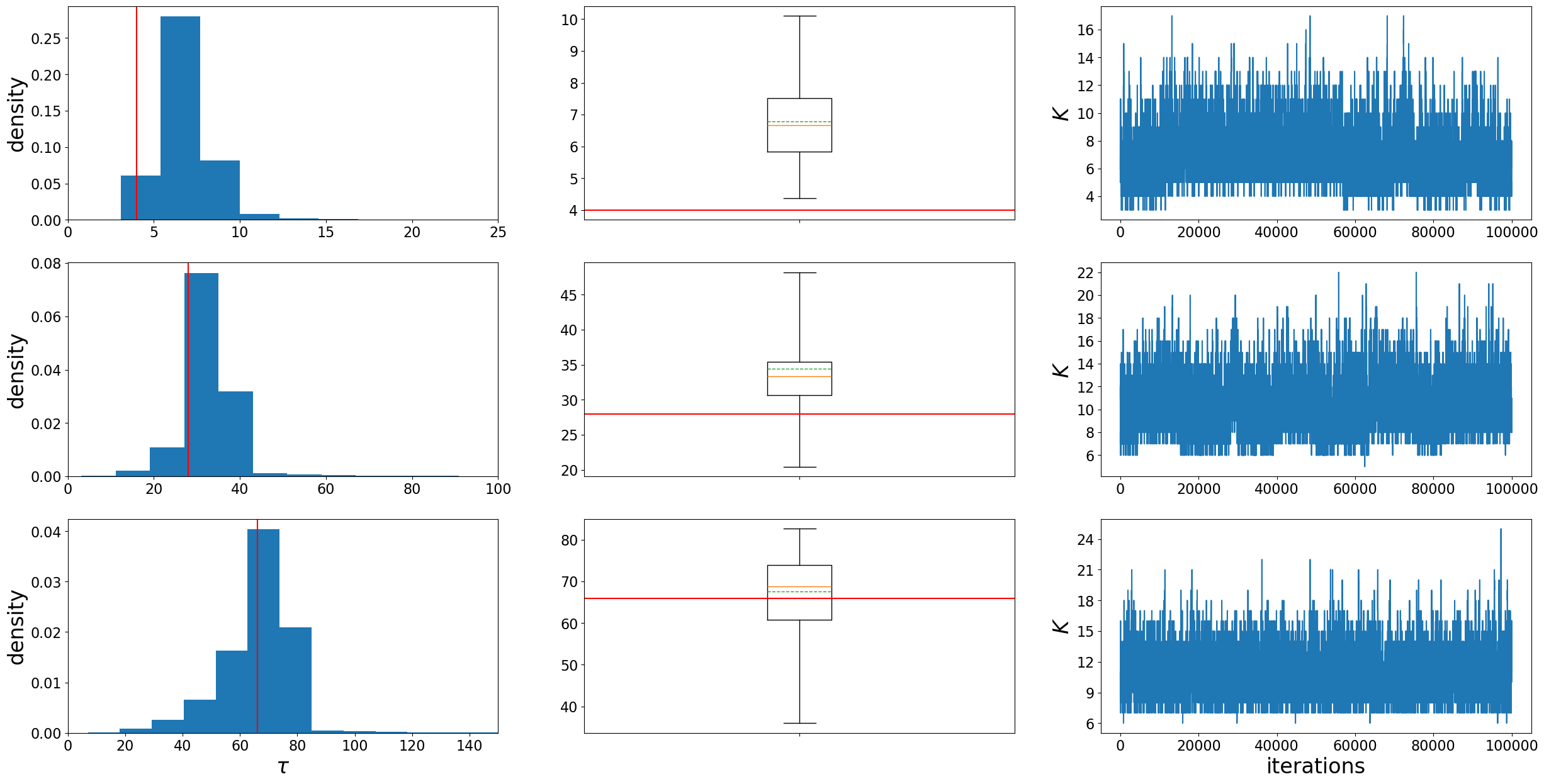}
    \caption{HDP on synthetic data with $\tau_{1}$ estimated via Monte Carlo. In red the true $\tau_1$. For the box plots: in orange the median, in dashed green the mean, and the whiskers show 95\% credible intervals.}
    \label{fig:synth_BNP_MC}
\end{figure}

\begin{figure}[httb!]
    \centering
    \includegraphics[width=1\textwidth]{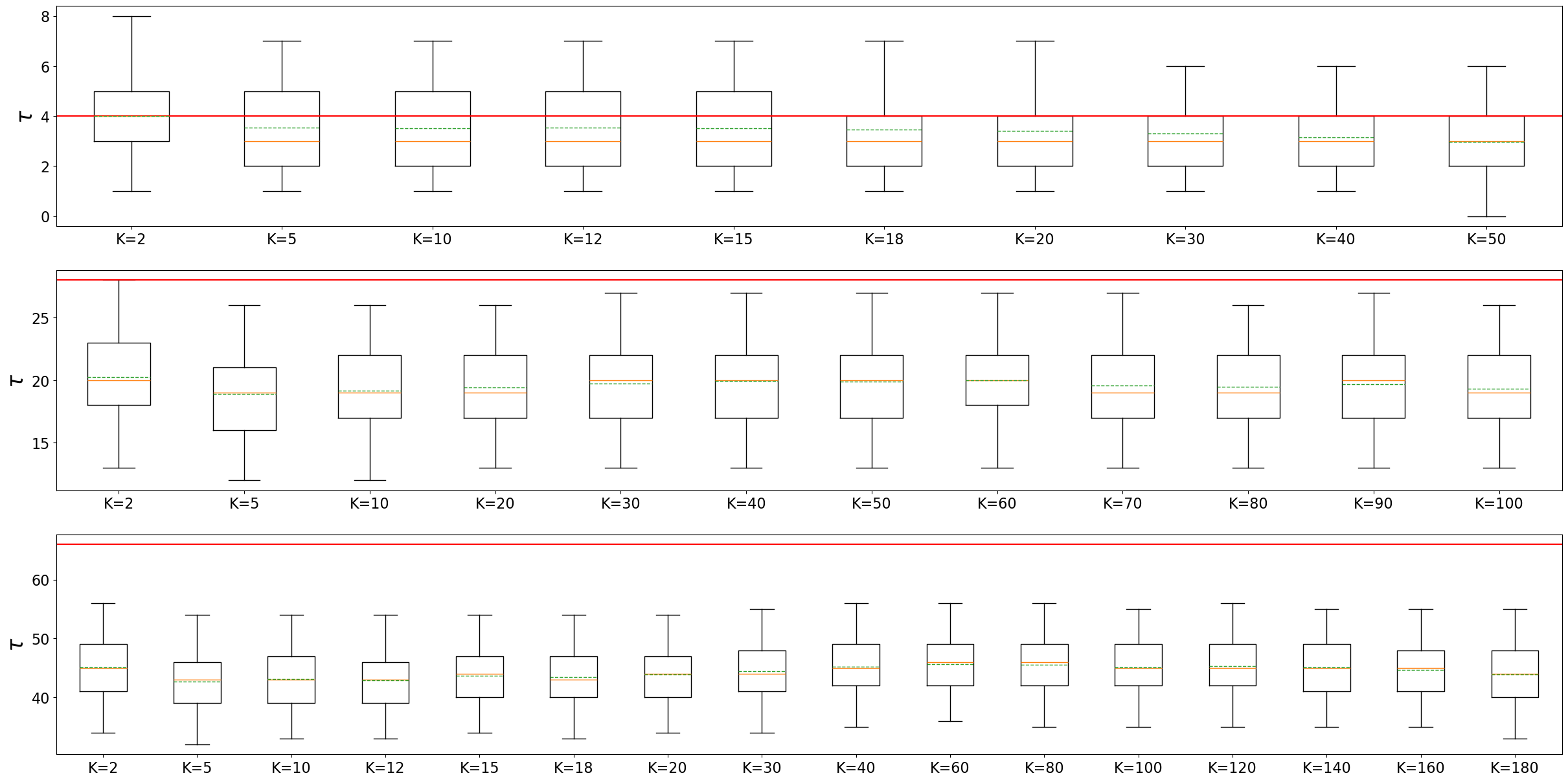}
    \caption{Estimates of $\tau_{1}$ with the parametric model for different values of $K$. In red the true $\tau_1$. For the box plots: in orange the median, in dashed green the mean, and the whiskers show 95\% credible intervals.}
    \label{fig:synth_MVR_sampling}
\end{figure}

Figures \ref{fig:synth_BNP_sampling}-\ref{fig:synth_BNP_MC}
display the histogram estimates (first column) and box plots (second column) of $\tau_{1}$ and the trace plots of the number of mixture components $K_{n}$ (third column) for the HDP model, using population and Monte Carlo sampling, respectively. Different rows in the figures correspond to different sample sizes. The non-parametric model performs well on synthetic data and is capable of recovering the true value of $\tau_{1}$, which is $4,28,66$ respectively and falls within 95\% credible intervals for all three sample sizes.
Moreover, the algorithms with population sampling and Monte Carlo sampling seem to have comparable performances, with the Monte Carlo approximation narrowing the credible intervals. In view of the computational time gain, the Monte Carlo sampling approach seems preferable, and we have focused on that in the real-data example.
Also, in Figure \ref{fig:synth_MVR_sampling}, the results of the parametric model are shown, from which it seems the model might slightly underestimate $\tau_{1}$. The overall posterior mean point estimates and standard deviations of $\tau_{1}$ are summarized in Table \ref{tab:synthetic_data} together with the computational times to run the algorithms.  

\begin{table}[httb!]
    \centering
    \begin{tabular}{ccccc}
    \hline
     Algorithm type  & $n$     & True $\tau_{1}$ & Est. $\tau_{1}$   & Comp. time (hours, max = 12h) \\
     HDP sampling    & 1000  & 4    & 6.66+/-2.42   & 12.01                  \\
     HDP sampling    & 5000  & 28   & 34.92+/-23.63 & 12.01                  \\
     HDP sampling    & 10000 & 66   & 67.33+/-25.42 & 12.0                   \\
     HDP Monte Carlo & 1000  & 4    & 6.79+/-1.55   & 1.49                   \\
     HDP Monte Carlo & 5000  & 28   & 34.48+/-20.86 & 3.13                   \\
     HDP Monte Carlo & 10000 & 66   & 67.64+/-24.78 & 5.35                   \\
     Parametric K=2         & 1000  & 4    & 4.01+/-1.68   & 5.39                   \\
     Parametric K=2         & 5000  & 28   & 20.26+/-3.67  & 5.95                   \\
     Parametric K=60        & 10000 & 66   & 45.63+/-5.25  & 6.89                   \\
    \hline
    \end{tabular}
    \caption{$\tau_{1}$ estimates of the HDP model under sampling and Monte Carlo and of the parametric model with the best choice of $K$ (the closest posterior mean to the true $\tau_1$).}
    \label{tab:synthetic_data}
\end{table}

\subsection{Real data} \label{sec:exp.real}

\begin{table}[httb!]
    \centering
    \resizebox{\textwidth}{!}{
    \begin{tabular}{l|rrrrrrrrrrr}
    EDUC & 0 & 1 & 2 & 3 & 4 & 5 & 6 & 7 & 8 & 9 & 10 \\
    \hline
    AGE &  &  &  &  &  &  &  &  &  &  &  \\
    0 & 69386 & 77787 & 51360 & 2180 & 298 & 50 & 1 & 0 & 0 & 0 & 0 \\
    1 & 36 & 12 & 5016 & 6070 & 1867 & 176 & 115 & 0 & 0 & 0 & 0 \\
    2 & 44 & 6 & 933 & 4159 & 5774 & 1807 & 330 & 33 & 7 & 0 & 0 \\
    3 & 51 & 17 & 448 & 963 & 4157 & 5565 & 2178 & 73 & 3 & 0 & 0 \\
    4 & 754 & 268 & 2085 & 2011 & 3189 & 6908 & 34299 & 19418 & 4507 & 8339 & 762 \\
    5 & 1673 & 507 & 4022 & 2857 & 3639 & 3966 & 50900 & 18474 & 13830 & 29715 & 15523 \\
    6 & 2846 & 944 & 6466 & 3943 & 4822 & 5220 & 84127 & 27168 & 21237 & 35888 & 27232 \\
    7 & 3444 & 1764 & 10976 & 4872 & 5860 & 5098 & 73838 & 17735 & 9791 & 21102 & 23182 \\
    8 & 2402 & 1306 & 11635 & 3765 & 4428 & 3490 & 38307 & 6884 & 1973 & 6629 & 6154 \\
    \hline
    \end{tabular}
    }
    \caption{Cross table between AGE and EDUC.}
    \label{tab:cross_table_age_educ}
\end{table}

We test the performances of the HDP model on a real dataset from the 5\% public use microdata sample (PUMS) of the 2000 U. S. census for the state of New York \citep{Rug24}. It contains information about the following 10 categorical variables, observed for a population of $953076$ individuals: ownership of dwelling (OWNERSHIP: 3 levels), mortgage status (MORTGAGE: 4 levels), age bands (AGE: 9 levels), sex (SEX: 2 levels), marital status (MARST: 6 levels), race identification band (RACESING: 5 levels), education level (EDUC: 11 levels), employment status (EMPSTAT: 4 levels), work disability (DISABWRK: 3 levels), and veteran status (VETSTAT: 3 levels). 

This data results in a contingency table of $2566080$ cells in total, many of which can considered as structural zeros. For example, from Table \ref{tab:cross_table_age_educ}, which cross-classifies age and education, we can see that there are some obvious structural zeros. These are due to the impossibility of some values of the categorical variables to coexist, e.g. age below 14 (level 1) with the highest level of education (level 11). Following \cite{Man14}, we recover $60$ overlapping marginal conditions, resulting in $557$ disjoint marginal conditions and representing $2317030$ cells of our contingency table (approx. 90$\%$). 

% Note that we obtained slightly fewer dijoint marginal conditions compared to \cite{Man14}, this is due to a different processing ordering of the overlapping marginal conditions representing the same number of cells, e.g. if marginal conditions 1 and marginal conditions 2 represent the same number of cells processing 1 and then 2 or the reverse will produce different outcomes.

As an initial demonstration of the performance of the non-parametric model, we first pre-processed the data to remove the majority of structural zeros and then implemented the algorithm from Section \ref{sec:sampling}. Specifically, we have removed all the individuals that were younger than $18$ and we dropped the categorical variables OWNERSHP and MORTGAGE. This results in a dataset with $712174$ individuals and a significantly smaller contingency table of $39600$ cells, with many fewer zero cells. For example, now the dataset contains only rows from categories 4 to 8 of Table \ref{tab:cross_table_age_educ}, hence reducing substantially the number of empty cells.
%For example, Table \ref{tab:cross_tabel_nozeros} shows the result of this pre-processing for the cross table between AGE and EDUC.

\begin{figure}[httb!]
    \centering
    \includegraphics[width=0.8\textwidth]{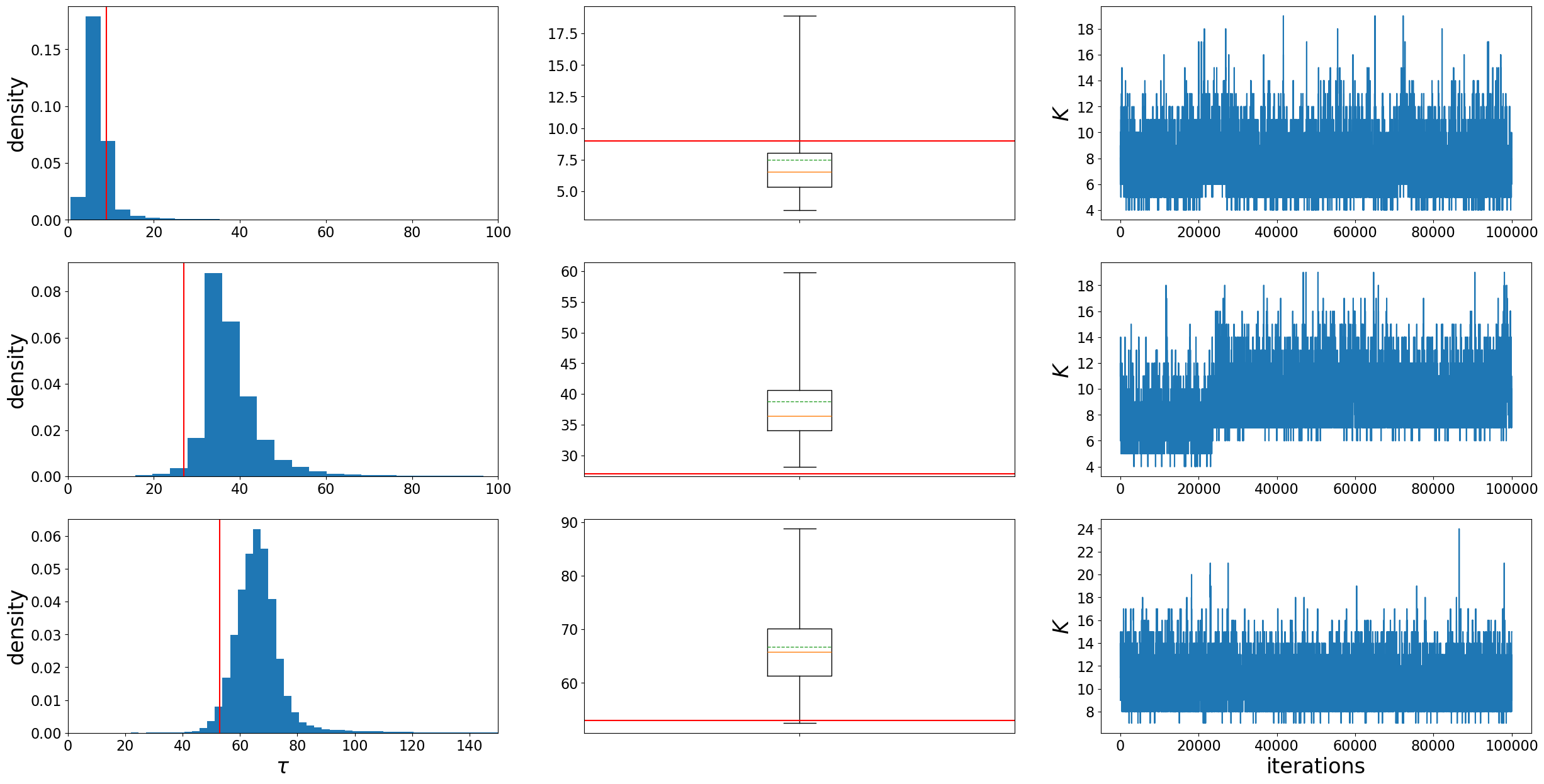}
    \caption{Estimates of $\tau_{1}$ with HDP, algorithm from Section \ref{sec:sampling}, on real data, after pre-processing to remove structural zeros. In red the true $\tau_1$. For the box plots: in orange the median, in dashed green the mean, and the whiskers show 95\% credible intervals.}
    \label{fig:NY_BNP_sampling}
\end{figure}

After drawing three samples of sizes $1000,5000,10000$, the MCMC of Section \ref{sec:sampling} was run for 300k iterations, with the first 200k iterations discarded as a burn-in.  Figure \ref{fig:NY_BNP_sampling} displays the posterior histogram and box plot of $\tau_{1}$, together with the trace plots of the number of mixture components $K_{n}$. Moreover, the first three rows of  Table \ref{tab:NY_NY_SZ} summarize point estimate and credible intervals of $\tau_{1}$. The true value of $\tau_{1}$ is mostly within the 95$\%$ credible interval. Note that the slight deterioration in performance compared to the synthetic data example might also be due to the presence of some additional structural zeros that have not been completely removed with the pre-processing step.

\begin{figure}[httb!]
    \centering
    \includegraphics[width=0.8\textwidth]{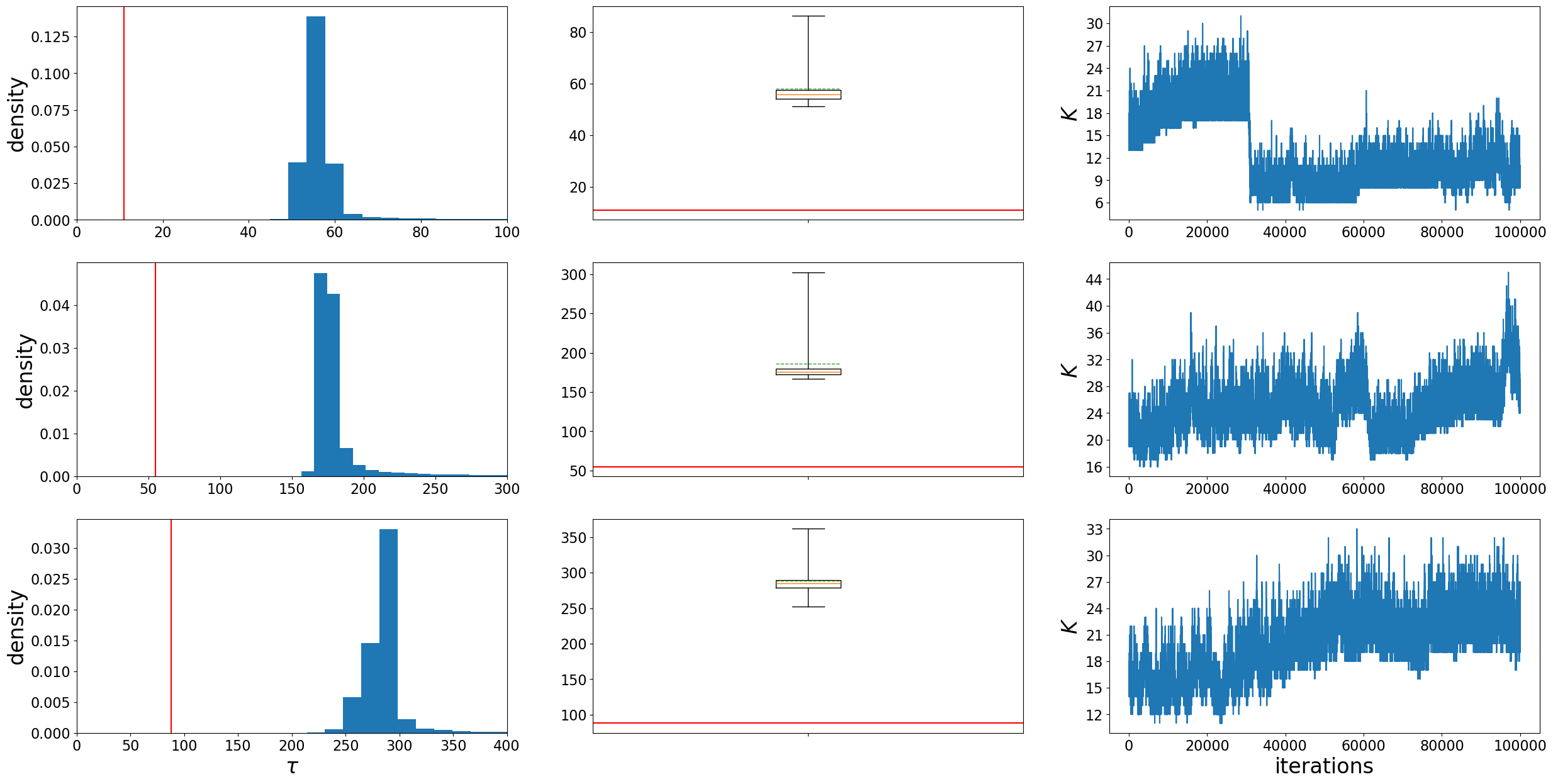}
    \caption{Estimates of $\tau_{1}$ with HDP, algorithm from Section \ref{sec:sampling}, on real data, without pre-processing to remove structural zeros. In red the true $\tau_1$. For the box plots: in orange the median, in dashed green the mean, and the whiskers show 95\% credible intervals.}
    \label{fig:NY_BNP_sampling_SZ}
\end{figure}

We now consider the raw data, in which structural zeros have not been removed. In order to show how much they can deteriorate the performances of the algorithm, we first run the algorithm of Section \ref{sec:sampling}, which does not model structural zeros. The results are displayed in Figure \ref{fig:NY_BNP_sampling_SZ} and rows 4-6 of Table \ref{tab:NY_NY_SZ}. Even with $300$k iterations, the Markov chain fails to properly converge and estimates $\tau_{1}$ to be very far from the true values. Indeed, the algorithm significantly overestimates $\tau$, as sample unique values have now a very low posterior probability of being sampled again. This is because most of the posterior probability mass is now assigned to the structural zeros cells, and much larger sample sizes are probably needed to wash out the effect of the prior and shrink this probability to zero.

\begin{table}[httb!]
    \centering
    \begin{tabular}{ccccc}
    \hline
     Algorithm type     & $n$     & True $\tau$ & Est. $\tau$    & Structural zeros \\
     BNP Monte Carlo    & 1000  & 9    & 7.49+/-5.32    & False            \\
     BNP Monte Carlo    & 5000  & 27   & 38.8+/-12.0    & False            \\
     BNP Monte Carlo    & 10000 & 53   & 66.72+/-10.85  & False            \\
     BNP Monte Carlo    & 1000  & 11   & 57.98+/-13.25  & True             \\
     BNP Monte Carlo    & 5000  & 55   & 186.06+/-46.67 & True             \\
     BNP Monte Carlo    & 10000 & 88   & 288.7+/-47.15  & True             \\
    SZ BNP Monte Carlo & 1000  & 11   & 10.81+/-1.66   & True             \\
     SZ BNP Monte Carlo & 5000  & 55   & 56.16+/-3.78   & True             \\
     SZ BNP Monte Carlo & 10000 & 88   & 189.46+/-5.76  & True             \\
                        &         &             & 101.98+/-5.27  & True             \\
    \hline
    \end{tabular}
    \caption{Real data with and without structural zeros (SZ). The HDP has been launched with and without the adjustment for structural zeros.}
    \label{tab:NY_NY_SZ}
\end{table}

\begin{figure}[httb!]
\centering
\begin{subfigure}
    \centering
    \includegraphics[width=0.8\textwidth]{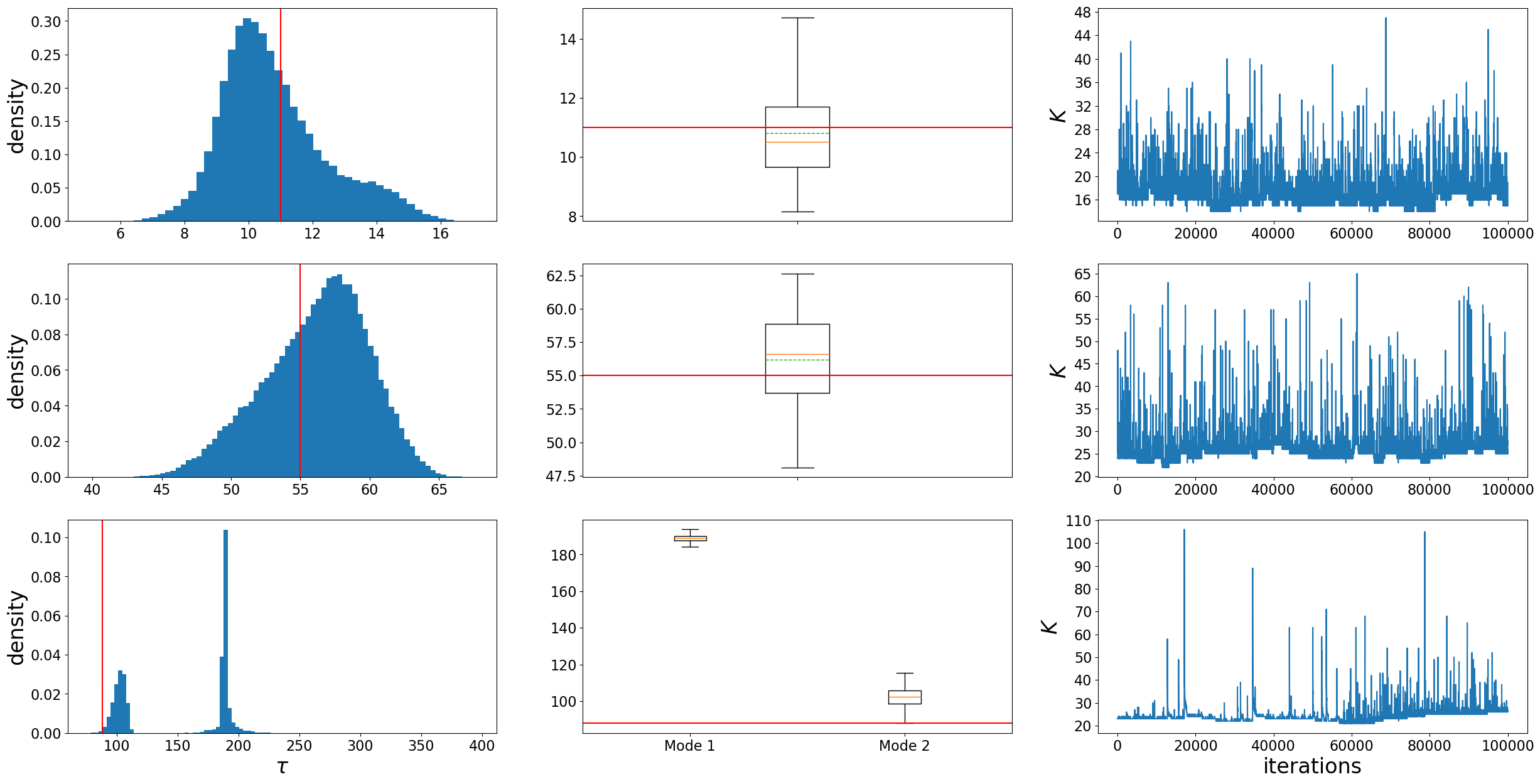}
\end{subfigure}
\begin{subfigure}
    \centering
    \includegraphics[width=0.8\textwidth]{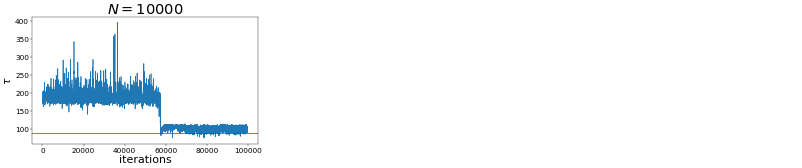}
\end{subfigure}
    \caption{Estimates of $\tau_{1}$ with HDP, algorithm from Section \ref{sec:mcmc_struc}, on real data, without pre-processing to remove structural zeros. The last row reports the trace plot for $N=10000$. In red the true $\tau_1$. For the box plots: in orange the median, in dashed green the mean, and the whiskers show 95\% credible intervals.}
    \label{fig:NY_BNP_sampling_SZ_SZ}
\end{figure}

Finally, we run the algorithm of \ref{sec:mcmc_struc}, which accounts for the presence of structural zeros in the data. Figure \ref{fig:NY_BNP_sampling_SZ_SZ} and the last three rows of Table \ref{tab:NY_NY_SZ} show the results of 100K iterations, obtained after the burn-in period of 100k, for three samples of sizes 1k, 5k, and 10k. The first two rows of Figure \ref{fig:NY_BNP_sampling_SZ_SZ}  display the histogram estimators and box plots of $\tau_{1}$ and the trace plots of $K_{n}$, for the samples  1k and 5k. From the plots, we can see that, for these samples, the MCMC has converged to stationary and the estimates $\tau_{1}$ are good, with the true value being within the 95$\%$ posterior credible intervals. However, we should warn that for some samples, the posterior distribution can become multi-modal. This is the case for the chosen sample of size 10k, third row in Figure \ref{fig:NY_BNP_sampling_SZ_SZ}. It is indeed evident that the posterior is bimodal. Due to the initialization chain, the chain spends many iterations in the sub-optimal mode, before jumping to the best mode. Estimates obtained by discarding more iterations of burn-in, or running for Markov chain longer, produce reasonable estimates of $\tau_{1}$.

\section{Discussion}

In this work, we have proposed a Bayesian non-parametric approach, based on hierarchical modelling and which generalizes parametric mixed membership models, to estimate measures of disclosure risk. The proposed approach does not have any tuning parameters and performs well in the experiments, even with samples as small as 1\% of the entire population. Also, the methodology can be extended to account for the presence of many structural zeros in the data, through a data augmentation scheme. Moreover, fast Monte Carlo approximation schemes have been suggested, which can reduce the computational cost of running the algorithms dramatically, hence making the approach applicable also in the presence of large population sizes $N$.

In terms of improvements, we have shown in the experiment section how the posterior distribution of the augmented model, accounting for structural zeros, can become multi-modal, depending on the observed sample. If the MCMC algorithm is initialized poorly and not run long enough, it can get stuck in a sub-optimal mode, hence producing misleading estimates of $\tau_{1}$. We recommend therefore that the practitioner tries to start the algorithm from different initial values of the parameters, in order to detect whether any multi-modality is present. If so, the MCMC should be run for many iterations to properly explore the parameter space and obtain accurate estimates. As a direction for improvement, it would be useful to define an automatic approach to detect multi-modality in the posterior distribution of mixed membership models and, if so, a way of properly handling it, like for example by restarting the algorithm from different initial values, choosing a good way to initialize the algorithm using the observed sample, or trying to improve step sizes and movement directions of the chain, to facilitate jumps from sub-optimal to optimal modes.

\bibliographystyle{name}
\bibliography{bibliography.bib}

\appendix

\section{Parametric Mixed Membership Model} \label{sec:ModelMan}

\emph{Mixed Membership models} are generalizations of mixture models to model multiple groups of observations. They assume $K$ extreme profiles (alias mixture components), having weights in the population regulated by a $K$-dimensional probability vector $\pmb{g}_{0}$. Within each group, some heterogeneity from the common proportions $\pmb{g}_{0}$ is allowed, by introducing a group-specific partial affiliation vector $\pmb{g}_{i}$. In the model used in \cite{Man12}, the $i$-th group of observations corresponds to the $J$ observations of key variables of the $i$-th individual. The model can be summarized as follows,
\begin{align} \label{eq:ModelMan}
    X_{i,j}|\mathbf{g}_{i},\{\theta^{\left(k\right)}\}_{k=1}^{K} & \sim \sum_{k=1}^{K}g_{i,k}\theta^{\left(k\right)}_{j,\cdot} \ \ \ \ & i=1,\ldots,n, \ \ j=1,\ldots,J \nonumber \\
    \mathbf{g}_{i} | \mathbf{g_{0}} & \sim \text{Dir}\left( \mathbf{g_{0}}\right) \ \ \ \ & i=1,\ldots,n \nonumber  \\ 
    \mathbf{g_{0}} & \sim \text{Dir}\left( \mathbb{I}_{K}\right)\cdot \alpha_{0} & \\
    \alpha_{0} & \sim \text{Ga}\left(2,1\right) &  \nonumber \\ 
    \theta_{j,\cdot}^{\left(k\right)} & \sim \text{Dir}\left( \mathbb{I}_{n_{j}}\right) \ \ \ & k=1,\ldots,K, \ \ \ j=1,\ldots,J \nonumber
\end{align}
where $\text{Ga}$ denotes a Gamma distribution, $\text{Dir}$ a Dirichlet distribution and $\mathbb{I}_{n_{j}}$ a vector of dimension $n_{j}$ with all entries equal to $1$. Moreover, for each $k\in\{1,\ldots,K\}$, $\theta^{\left(k\right)}$ is an array with $J$ rows, whose $j$-th row, denoted $\theta_{j,\cdot}^{\left(k\right)}$, is a probability vector of dimension $n_{j}$, the number of categories of variable $j$.

For computational reasons, it is useful to do data augmentation introducing classification variables $Z_{i,j}$ for the mixture components, in the first layer of the hierarchy. This is achieved by replacing the first line of \eqref{eq:ModelMan} with 
\begin{align*}
    X_{i,j}|\left(Z_{i,j}=k\right) & \sim \theta^{\left(k\right)}_{j,\cdot} \\
  \mathbb{P}\left( Z_{i,j}= k |\mathbf{g}_{i}\right) & = g_{i,k} \ \ \ \ k=1,\ldots,K  
\end{align*}
for all $i=1,\ldots,n$ and $ j=1,\ldots,J$.

%%%%%%%%%%%%%%%%%%%%%%%%%%%%%
%%%%%%%%%%%%%%%%%%%%%%%%%%%%%
%%%%%%%%%%%%%%%%%%%%%%%%%%%%%

\section{Monte Carlo approximation of $\tau_{1}$} \label{sec:MonteCarlo}

Let us recall that $\mathcal{C}:=\bigtimes_{j=1}^{J}\{1\ldots,n_{j}\}$ denotes the state space of the observations, and $(f_{1},\ldots,f_{|\mathcal{C}|})$ the vector of observed frequencies in the sample. Moreover, let us recall that the risk measure $\tau_{1}$ is defined as
\begin{equation} \label{riskindeces2}
        \tau_{1}=\sum_{c\in \mathcal{C}:f_{c}=1}r_{1c}, \ \ \ \text{where} \ \ \  r_{1c}=\mathbb{P}\left(F_{c}=1|f_{c}=1\right). 
\end{equation}
$\tau_{1}$ can also be rewritten as
\begin{equation*}
    \tau_{1} = \sum_{c \in \mathcal{C}} \mathbb{I}\left( \sum_{i=1}^{n} \mathbb{I}\left(X_{i}=c\right)=1\right) \mathbb{I}\left(\sum_{i=n+1}^{N} \mathbb{I}\left(X_{i}=c\right)=0\right).
\end{equation*}

Given a sample $X_{1:n}:=\left(X_{1},\ldots,X_{n}\right)$, and denoting by $\tilde{\mathcal{C}}_{X_{1:n}}$ the set $\tilde{\mathcal{C}}_{X_{1:n}}:=\{c\in \mathcal{C}: \sum_{c\in\mathcal{C}} \mathbb{I}\left(\sum_{i=1}^{n}\mathbb{I}\left(X_{i}=c\right)=1\right)\}$ the set of combinations appearing with frequency $1$ in the sample, the expected value of $\tau_{1}$ is
\begin{equation*}
   \mathbb{E}\left( \tau_{1}| X_{1:n}\right)= \sum_{c\in\tilde{\mathcal{C}}_{X_{1:n}}} \mathbb{P}\left( \sum_{i=n+1}^{N}\mathbb{I}\left(X_{i}=c\right)=0 |X_{1:n}\right).
\end{equation*}
This can be estimated within the MCMC sampler as follows, by conditioning on $G_{0}$ (alias $\left(g_{0,k}\right)$ and $\left(\theta^{\left(k\right)}\right)$), 
\begin{align} \label{eq:appr.tau}
     \mathbb{E}&\left( \tau_{1}|  X_{1:n},G_{0},\alpha_{0}\right) = \sum_{c\in\tilde{\mathcal{C}}_{X_{1:n}}} \mathbb{P}\left( \sum_{i=n+1}^{N}\mathbb{I}\left(X_{i}=c\right)=0 |X_{1:n},G_{0},\alpha_{0}\right) \nonumber \\
     & = \sum_{c\in\tilde{\mathcal{C}}_{X_{1:n}}} \mathbb{P}\left( \sum_{i=n+1}^{N}\mathbb{I}\left(X_{i}=c\right)=0 |G_{0},\alpha_{0}\right) 
       = \sum_{c\in\tilde{\mathcal{C}}_{X_{1:n}}} \mathbb{P}\left( \cap_{i=n+1}^{N}\{X_{i}\neq c\} |G_{0},\alpha_{0}\right) \nonumber \\
       & = \sum_{c\in\tilde{\mathcal{C}}_{X_{1:n}}} \prod_{i=n+1}^{N} \mathbb{P}\left( \{X_{i}\neq c\} |G_{0},\alpha_{0}\right)  = \sum_{c\in\tilde{\mathcal{C}}_{X_{1:n}}}  \mathbb{P}\left( \{X_{n+1}\neq c\} |G_{0},\alpha_{0}\right)^{N-n} \nonumber \\
         & = \sum_{c\in\tilde{\mathcal{C}}_{X_{1:n}}}  \left(1- \mathbb{P}\left( \{X_{n+1}= c \} |G_{0},\alpha_{0}\right)\right)^{N-n}
\end{align}
where the second and fourth equalities follow because 
individuals are conditionally independent given $G_{0}$, while the fifth equality is because they are also conditionally identically distributed. Finally, $\mathbb{P}\left( \{X_{n+1}= c \} |G_{0},\alpha_{0}\right)$ can be approximated by writing 
\begin{align} \label{eq:MC_prob}
    \mathbb{P}\left( \{X_{n+1}= c \} |G_{0},\alpha_{0}\right) & = \int \mathbb{P}\left( \{X_{n+1}= c \}  |G_{i},\alpha_{0}\right)\mathbb{P}\left(G_{i}|G_{0},\alpha\right) dP\left(G_{i}\right) \nonumber \\
    & = \int \left(\prod_{j=1}^{J} \left(\sum_{k=1}^{K_{n}} g_{i,k}  \theta^{\left(k\right)}_{j,c_j}  + g_{i,0} \frac{1}{n_{j}}\right) \right) \text{Dir}\left(\left(g_{i,k}\right) | \left(g_{0,k}\right) \right)   d\left(g_{i,k}\right)
\end{align}
Estimation of this integral can quickly be computed by Monte Carlo by sampling from $\text{Dir}\left(\left(g_{i,k}\right) | \left(g_{0,k}\right)\right)$, the density of the Dirichlet distribution from equation (5) of the main paper. 

\subsubsection*{Algorithm with structural zeros:}
Following similar steps, formula \ref{eq:appr.tau} can be extended to the algorithm with structural zeros, to use the approximation
\begin{equation} \label{eq:appr.tau.struc}
\tau_{1} \approx \sum_{c\in\tilde{\mathcal{C}}_{X_{1:n}}}  \left(1- \mathbb{P}\left( \{X_{n+1}= c \} |G_{0},\alpha_{0}\right)\right)^{(N-n)\frac{1}{1-p_{0}}}
\end{equation}
The exponent $(N-n)\frac{1}{1-p_{0}}$ follows from the fact that in the data augmentation algorithm with structural zeros, $n_{0}$ (the total number of augmented observations among all structural zeros cells) is distributed according to a Negative Binomial distribution of parameters $n$ and $1-p_{0}$, from the properties of the Negative Multinomial, where $p_{0}=\sum_{c=1}^{C}p_{c}$, sampled in step 7 of the algorithm, is the probability of falling into any of the $C$ structural zero cells. The expected value of $n_{0}$ is therefore $n\frac{p_{0}}{1-p_{0}}$. Similarly, if we denote by $N_{0}$ the ideal number of individuals in the entire population (of size $N+N_{0}$) belonging to structural zeros cells, its distribution is a Negative Binomial of parameters $N$ and $1-p_{0}$, and its mean is $N\frac{p_{0}}{1-p_{0}}$. \\
Replacing the boundaries of the sum $\sum_{i=n+1}^{N}$ in \eqref{eq:appr.tau} with $\sum_{i=n+n_{0}+1}^{N+N_{0}}$, following the same steps and using $\mathbb{E}((N+N_{0})-(n+n_{0}))=(N-n)\frac{1}{1-p_{0}}$, we obtain the approximation  \eqref{eq:appr.tau.struc}.

%%%%%%%%%%%%%%%%%%%%%%%%%%%%%%
%%%%%%%%%%%%%%%%%%%%%%%%%%%%%%
%%%%%%%%%%%%%%%%%%%%%%%%%%%%%%

\section{A fast approximation of Step 9 for the MCMC with structural zeros (Section 4.1 of main paper)}  \label{sec:step9}

In Section 4.1 of the main paper, we have described an algorithm to perform posterior inference in the presence of structural zeros. This algorithm relies on a data scheme, in which some additional observations, taking values in the structural zero cells and denoted $X_{\left(n+1\right):\left(n+n_{0}\right)}$, are generated within the sampler. However, 
when the structural zeros account for the majority of cells, the number of augmented observations can become very large. This is because the probabilities $(p_{1},\ldots,p_{C})$ and counts $(n_{1},\ldots,n_{C})$ from steps 7 and 8 (Section 4.1 of the main paper) can become very large. This implies that, at step 9, many variables need to be simulated, and this slows down the algorithm significantly. In this section, we describe an approximation of step 9 that reduces the computational cost dramatically and produces similar estimates of $\tau_{1}$. 

We should notice that in steps 1-6, the only information from $X_{\left(n+1\right):\left(n+n_{0}\right)}$ and $Z_{\left(n+1\right):\left(n+n_{0}\right)}$ that is needed to update the common parameters are:  1) $\sum_{i=n+1}^{n+n_{0}}\mathbb{I}\left(Z_{ij}=k,X_{ij}=c\right)$, for every $j$, $k$, and $c\leq n_{j}$, in order to update $\theta^{\left(k\right)}$ at step 5; 2)  $\sum_{i=n+1}^{n+n_{0}} m_{k1}$, for all $k$, needed in step 3 to update $\left(g_{0,k}\right)$. A fast approximation of both quantities can be obtained by taking their expectations and approximating them with Monte Carlo. We draw $T$ i.i.d. values of $\left(g_{t,k}\right)$ from formula (5) of the main paper, and approximate the two quantities as follows. 

Regarding item 1,  we can use the approximation 
\begin{align*}
\mathbb{E}&\left(\sum_{i=n+1}^{n+n_{0}} \mathbb{I}\left(Z_{ij}=k,X_{ij}=x\right)|G_{0},\pmb{n}_{c}\right)  = \mathbb{E}\left(\sum_{c=1}^{C}\sum_{i}\mathbb{I}\left(Z_{ij}=k, X_{i,j} =x \right)|G_{0},\pmb{n}_{c}\right) \\    
&=\sum_{c=1}^{C}\sum_{i}\mathbb{P}\left(Z_{ij}=k, X_{i,j} = x |G_{0}, X_{i}\in \mu_{c} \right)  \\
&=\sum_{c=1}^{C}n_{c}\mathbb{P}\left(Z_{ij}=k, X_{i,j} = x |G_{0}, X_{i}\in \mu_{c} \right)  \\
&=\sum_{c=1}^{C}n_{c}\mathbb{P}\left(Z_{ij}=k| X_{i,j} = x ,G_{0}\right) \frac{1}{n_{j}}\mathbb{I}\left(x\in \mu_{c,j}\right)  \\
   &  \approx \sum_{c=1}^{C}n_{c} \frac{1}{T}\sum_{t=1}^{T} \frac{g_{t,k} \theta^{\left(k\right)}_{j,x} }{\sum_{k=1}^{K_{n}} g_{t,k}\theta^{(k)}_{j,x} + g_{t,0}\frac{1}{n_{j}}} 
 \frac{1}{n_{j}}\mathbb{I}\left(x\in \mu_{c,j}\right)
\end{align*}
where the sum in $i$ runs from $i=n + \sum_{l=1}^{c-1}n_{l}+1$ to $n + \sum_{l=1}^{c-1}n_{l}+n_{c}$, i.e. over the $n_{c}$ $X_{i}$ assigned to the $c$-th marginal constraint. Also, notice that we write $x \in \mu_{c,j}$ (and not $x = \mu_{c,j}$) so that the indicator is equal to one if either $\mu_{c,j}=*$ or $x$ is equal to specific value fixed by the $c$-th marginal constraint in the $j$-th variable.

Regarding item 2,  we first approximate for $t\in \{ 1,\ldots,T \}$ and $c \in \{1,\ldots,C\}$, the quantity $\sum_{j=1}^{J} \mathbb{I}\left(Z_{ij}=k,X_{ij}=\mu_{c,j}\right)$ by its expectation given the event $X_{i}\in \mu_{c}$ (hence the superscript $c$ in $n_{i\cdot k}^{t,c}$)  and given $g_{t,k}$ (hence the superscript $t$ in $n_{i\cdot k}^{t,c}$), 

 \begin{align*}
   n_{i\cdot k}^{t,c}  & := \mathbb{E}\left(\sum_{j=1}^{J} \mathbb{I}\left(Z_{ij}=k,X_{ij}=\mu_{c,j}\right)|g_{t,k},X_{i}\in \mu_{c}\right) \\
   & = \mathbb{E}\left(\sum_{j=1}^{J} \mathbb{I}\left(Z_{ij}=k\right)|g_{t,k},X_{i}\in \mu_{c}\right)  =\sum_{j=1}^{J} \mathbb{P}\left(Z_{ij}=k|g_{t,k},X_{i}\in \mu_{c}\right)  \\
  &=\sum_{j=1}^{J}  
   \frac{ g_{t,k} \sum_{x\in \mu_{c,j}}\theta^{(k)}_{j,x}}{\sum_{k=1}^{K_{n}} g_{t,k}\sum_{x\in \mu_{c,j}}\theta^{(k)}_{j,x} + g_{t,0}\sum_{x\in \mu_{c,j}}\frac{1}{n_{j}}}  \\
   & =\sum_{j: \mu_{c,j} \neq *}  
   \frac{ g_{t,k} \theta^{(k)}_{j,\mu_{c,j}}}{\sum_{k=1}^{K_{n}} g_{t,k}\theta^{(k)}_{j,\mu_{c,j}} + g_{t,0}\frac{1}{n_{j}}} +  \sum_{j: \mu_{c,j} = *} g_{t,k}
 \end{align*}

 Given the $n_{i\cdot k}^{t,c}$ for $t \in \{1,\ldots,T\}$ and $c \in \{1,\ldots,C\}$, we approximate $m_{\cdot k}$ by
 %\begin{align*}
 %    \mathbb{E}&\left(  m_{\cdot,k} |g_{0,k}\right)  = \mathbb{E}\left( \sum_{i=n+1}^{n+n_{0}} m_{i,k} |g_{0,k}\right) = \mathbb{E}\left( \sum_{c=1}^{C}\sum_{i} m_{i,k} \mathbb{I}\left(X_{i} \in c\right) |g_{0,k}\right) \\
%    &  =  \sum_{c=1}^{C}\sum_{i}\mathbb{E}\left( m_{i,k} \mathbb{I}\left(X_{i} \in c\right) |g_{0,k}\right)  =   \sum_{c=1}^{C}n_{c}\mathbb{E}\left( m_{i,k} \mathbb{I}\left(X_{i} \in c\right) |g_{0,k}\right)   \\
%     & \approx  \sum_{c=1}^{C}n_{c} \frac{1}{T} \sum_{t=1}^{T} \mathbb{E}\left( m_{i,k} \mathbb{I}\left(X_{i} \in c\right) |g_{0,k}, n_{i\cdot k}^{t,c}\right) =  \sum_{c=1}^{C}n_{c} \frac{1}{T} \sum_{t=1}^{T} \sum_{l=1}^{n_{i\cdot k}^{t,c}} \frac{\alpha_{0}g_{0,k}}{\alpha_{0}g_{0,k}+l -1}
% \end{align*}
 \begin{align} \label{eq:tables}
     \mathbb{E}&\left(  m_{\cdot,k} |g_{0,k}\right)  = \mathbb{E}\left( \sum_{i=n+1}^{n+n_{0}} m_{i,k} |g_{0,k}\right) = \sum_{c=1}^{C}\sum_{i}\mathbb{E}\left(  m_{i,k} |g_{0,k}, X_{i}\in \mu_{c} \right) \nonumber \\
    &  =  \sum_{c=1}^{C}n_{c}\mathbb{E}\left(  m_{i,k} |g_{0,k}, X_{i}\in \mu_{c} \right)    \approx  \sum_{c=1}^{C}n_{c} \frac{1}{T} \sum_{t=1}^{T} \mathbb{E}\left( m_{i,k} |g_{0,k}, n_{i\cdot k}^{t,c}\right) \nonumber \\
    & =  \sum_{c=1}^{C}n_{c} \frac{1}{T} \sum_{t=1}^{T} \sum_{l=1}^{n_{i\cdot k}^{t,c}} \frac{\alpha_{0}g_{0,k}}{\alpha_{0}g_{0,k}+l -1} 
 \end{align}
where again the sum in $i$ runs from $i=n + \sum_{l=1}^{c-1}n_{l}+1$ to $n + \sum_{l=1}^{c-1}n_{l}+n_{c}$, i.e. over the $n_{c}$ $X_{i}$ assigned to the $c$-th marginal constraint. The last equality follows the expected number of tables in a Chinese Restaurant Process with $n_{i\cdot k}^{t,c}$ customers and parameter $\alpha_{0}g_{0,k}$.

In a similar fashion, we compute $m_{\cdot 0}$, the number of tables of variables in unobserved individuals $X_{\left(n+1\right):\left(n+n_{0}\right)}$ assigned to new mixtures, by first computing $n_{i\cdot 0}^{t,c}$ for each $t$ and $c$ as,
\begin{equation*}
    n_{i\cdot 0}^{t,c}  =\sum_{j: \mu_{c,j} \neq *}  
   \frac{ g_{t,k} \frac{1}{n_{j}}}{\sum_{k=1}^{K_{n}} g_{t,k}\theta^{(k)}_{j,\mu_{c,j}} + g_{t,0}\frac{1}{n_{j}}} +  \sum_{j: \mu_{c,j} = *} g_{t,k}
\end{equation*}
and then applying formula \eqref{eq:tables} to compute $m_{\cdot 0}$.
Finally, we replace step 5 in the algorithm of Section 4.1 of the main paper, to update $\left(g_{0,k}\right)$, with
    \begin{equation*} \text{Dir}\left(\alpha_{0}+m_{\cdot 0},m_{\cdot 1}, \ldots, m_{\cdot K_{n}}\right)
    \end{equation*}
and, in step 6 of the algorithm, when updating $\alpha_{0}$, we compute $m_{\cdot \cdot}=\sum_{k=0}^{K_{n}}m_{\cdot k}$, hence also including $m_{\cdot 0}$.  

\section{Notes on the implementation} \label{sec:implementation}

As briefly mentioned in the main paper, we have run experiments on GPUs, which are well-known to be more suited to parallel operations than serials. In this section, we comment on some of the implementation tricks that allowed us to gain the most out of the hardware and so get a considerable speed-up compared to the corresponding serial implementation. 

Consider the steps from the basic MCMC sampler to be used in the absence of structural zeros.

\begin{enumerate}
    \item This step can be implemented sequentially in $i$ leading to a cost that is linear in the sample size. Another option is to assign all the $i$'s to the mixtures labels $0,1,\dots, K_n$ in parallel with $0$ being the ``new mixtures'' label. At this point, we can restrict on the $i$'s assigned to $0$ and repeat the previous step on the labels $0,K_n+1$, sequentially assigning the remaining $i$'s to either a new mixture $0$ or the novel mixture $K_n+1$. We can then keep repeating the previous until all the $i$'s have a label that is different from $0$. Depending on the probability of being assigned to a new mixture the computational cost is reduced dramatically.
    \item Here we are interested in the number of tables $m_{ik}$ resulting from a Chinese Restaurant Process with $n_{i \cdot k}$ customers and concentration parameter $\alpha_0 g_{0,k}$. Instead of running the algorithm sequentially and assigning each customer to an old or a new table, one can notice that $m_{ik}$ is a sum of Bernoulli's with probability parameter given by $\frac{\alpha_0 g_{0,k}}{\alpha_0 g_{0,k} +t -1}$, where $t$ is the number of iterations of the Chinese Restaurant Process. We can then sample $n_{i \cdot k}$ Bernoulli's in parallel and then count the successes to get our $m_{ik}$.
\end{enumerate}
The other steps of the algorithm are already parallel. Remark also that even the estimate of $\tau_1$ via sampling can be parallelised in $N$ as we just need to know if in the new sample there are any rows that are sample unique in the original sample. However, the computational cost remains considerably larger than the Monte Carlo approach.

About the algorithm in the presence of structural zeros, step 1-6 remain the same and the novel 7-8-9 steps are easily parallelisable using the aforementioned tricks.

\end{document}